\documentclass[%
 preprint,
 superscriptaddress,
 amsmath,amssymb,
 aps,
 prc,
]{revtex4-1}

\usepackage{graphicx}
\usepackage{dcolumn}
\usepackage{bm}
\usepackage{url}
\usepackage{hyperref}

\begin{document}

\preprint{LLNL-JRNL-744244}

\title{Lifetimes of low-lying excited states in $^{86}_{36}$Kr$_{50}$}

\author{J.~Henderson}
\email{henderson64@llnl.gov}
\affiliation{Lawrence Livermore National Laboratory, Livermore, CA 94550, USA}
\affiliation{TRIUMF, Vancouver BC Canada V6T 2A3}
\author{A.~Chester}
\affiliation{Department of Chemistry, Simon Fraser University, Burnaby, BC V5A 1S6, Canada}
\author{G.~C.~Ball}
\author{R.~Caballero-Folch}
\affiliation{TRIUMF, Vancouver BC Canada V6T 2A3}
\author{T.~Domingo}
\affiliation{Department of Chemistry, Simon Fraser University, Burnaby, BC V5A 1S6, Canada}
\author{T.~E.~Drake}
\affiliation{Department of Physics, University of Toronto, Toronto, Ontario M5S 1A7, Canada}
\author{L.~J.~Evitts}
\affiliation{TRIUMF, Vancouver BC Canada V6T 2A3}
\affiliation{{Department of Physics, University of Surrey, Guildford, GU2 7XH, United Kingdom}}
\author{G.~Hackman}
\affiliation{TRIUMF, Vancouver BC Canada V6T 2A3}
\author{S.~Hallam}
\altaffiliation[Present address: ]{Department of Physics, University of Surrey, Guildford, GU2 7XH, United Kingdom}
\affiliation{TRIUMF, Vancouver BC Canada V6T 2A3}
\affiliation{{Department of Physics, University of Surrey, Guildford, GU2 7XH, United Kingdom}}
\author{A.~B.~Garnsworthy}
\author{M.~Moukaddam}
\altaffiliation[Present address: ]{Department of Physics, University of Surrey, Guildford, GU2 7XH, United Kingdom}
\author{P.~Ruotsalainen}
\altaffiliation[Present address: ]{Department of Physics, University of Jyv\"{a}skyl\"{a}, FIN-40014 Finland}
\author{J.~Smallcombe}
\author{J.~K.~Smith}
\altaffiliation[Present address: ]{Physics Department, Reed College, Portland OR, 97202, USA}
\affiliation{TRIUMF, Vancouver BC Canada V6T 2A3}
\author{K.~Starosta}
\email{starosta@sfu.ca}
\affiliation{Department of Chemistry, Simon Fraser University, Burnaby, BC V5A 1S6, Canada}
\author{C.~E.~Svensson}
\affiliation{Department of Physics, University of Guelph, Guelph, ON N1G 2W1, Canada}
\author{J.~Williams}
\affiliation{Department of Chemistry, Simon Fraser University, Burnaby, BC V5A 1S6, Canada}

\date{\today}

\begin{abstract}
\edef\oldrightskip{\the\rightskip} 
\begin{description}
\rightskip\oldrightskip\relax
\item[Background] The evolution of nuclear magic numbers at extremes of isospin is a topic at the forefront of contemporary nuclear physics. $N=50$ is a prime example, with increasing experimental data coming to light on potentially doubly-magic $^{100}$Sn and $^{78}$Ni at the proton-rich and proton-deficient extremes, respectively.
\item[Purpose] Experimental discrepancies exist in the data for less exotic systems. In $^{86}$Kr the $B(E2;2^+_1\rightarrow0^+_1)$ value - a key indicator of shell evolution - has been experimentally determined by two different methodologies, with the results deviating by $3\sigma$. Here, we report on a new high-precision measurement of this value, as well as the first measured lifetimes and hence transition strengths for the $2^+_2$ and $3^-_{(2)}$ states in the nucleus.
\item[Methods] The Doppler-shift attenuation method was implemented using the TIGRESS gamma-ray spectrometer and TIGRESS integrated plunger (TIP) device. High-statistics Monte-Carlo simulations were utilized to extract lifetimes in accordance with state-of-the-art methodologies.
\item[Results] Lifetimes of $\tau(2^+_1)=336\pm4\text{(stat.)}\pm20\text{(sys.)}$~fs, $\tau(2^+_2)=263\pm9\text{(stat.)}\pm19\text{(sys.)}$~fs and $\tau(3^-_{(2)})=73\pm6\text{(stat.)}\pm32\text{(sys.)}$~fs were extracted. This yields a transition strength for the first-excited state of $B(E2;2^+_1\rightarrow0^+)=259\pm3\text{(stat.)}\pm16\text{(sys.)}$~e$^2$fm$^4$.
\item[Conclusions] The measured lifetime disagrees with the previous Doppler-shift attenuation method measurement by more than $3\sigma$, while agreeing well with a previous value extracted from Coulomb excitation. The newly extracted $B(E2;2^+_1\rightarrow0^+_1)$ value indicates a more sudden reduction in collectivity in the $N=50$ isotones approaching $Z=40$.
\end{description}
\end{abstract}

\pacs{Valid PACS appear here}
\maketitle


\section{Introduction}

Determining the evolution of the nuclear magic-numbers far from the line of $\beta$-stability is the subject of much experimental and theoretical effort. In particular, the advent of the next generation of radioactive ion beam facilities presents the promise of further study of this evolution at the extremes of isospin. So-called ``islands of inversion'' are now well established at $N=8$, $N=20$ and $N=28$~\cite{ref:Navin_00,ref:Thibault_75,ref:Bastin_07}, all of which are established major shell-closures at the line of stability, as well as the sub-shell closure at $N=40$~\cite{ref:Naimi_12}. The $N=50$ shell may also be expected to exhibit a similar breakdown of the nominal magic numbers (see e.g.~\cite{ref:Santamaria_15,ref:Nowacki_16}), incorporating the neutron-deficient (nominally) doubly-magic $^{100}$Sn and neutron-rich $^{78}$Ni. Key experimental observables in the mapping of nuclear shell evolution include excited $2^+_1$ state energies and $E2$ transition strengths ($B(E2;2^+\rightarrow0^+)$). In order to properly assess the evolution of nuclear shell closures the experimental determination of these observables at stability where ``normal'' configurations dominate is thus important.

Of the even-even $N=50$ isotones, $B(E2;2^+_1\rightarrow0^+_1)$ data is available for seven nuclides, from $^{80}$Zn through to $^{92}$Mo. A clear minimum is found for $^{90}$Zr, which has $Z=40$ corresponding to a sub-shell closure and behaves very much like a doubly-magic nucleus. Between $Z=30$ and $Z=38$, the $B(E2)$ values are relatively consistent, while the data for heavier isotones is limited to $^{92}$Mo, preventing any systematic comparison. The data for $^{86}_{36}$Kr is found to be inconsistent, with $B(E2)$ values determined from a lifetime measurement using the DSAM technique~\cite{ref:Mertzimekis_01} found to deviate from that determined using Coulomb excitation~\cite{ref:Cheng-Lie_81} at a $3\sigma$ level. In the event that the $B(E2)$ value extracted from DSAM in Ref.~\cite{ref:Mertzimekis_01} is correct, the degree of collectivity characterized by the $B(E2)$ decreases steadily from a maximum at $^{82}$Ge towards the minimum at $^{90}$Zr. If, on the other hand, the Coulomb-excitation $B(E2)$ of Ref.~\cite{ref:Cheng-Lie_81} is correct, the decrease is rather more pronounced from a near-maximum at $Z=36$ to a minimum at $Z=40$. A figure showing $N=50$ $B(E2;2^+_1\rightarrow0^+_1)$ systematics is shown in the discussion section of this paper. 

In the present work, we therefore undertook to remeasure the lifetime of the $2^+_1$ state in $^{86}$Kr using the DSAM technique, following population by unsafe Coulomb-excitation. The major benefit of this population mechanism is that Coulomb excitation cross-sections reduce as one requires more excitation energy or further steps of excitation, greatly reducing the impact of feeding from higher lying states on the lifetimes measured for the states of interest. A benefit for the present measurement over that of Ref.~\cite{ref:Mertzimekis_01} is the use of a $4\pi$ HPGe array, greatly enhancing sensitivity to near-degenerate states which might only be resolved at either forward- or backward angles. The analysis of this work is very similar to that described in Ref.~\cite{ref:Chester_18} but utilizes the DSAM technique rather than the recoil-distance method.

\section{Experimental details}

\begin{figure}
\centerline{\includegraphics[width=\linewidth]{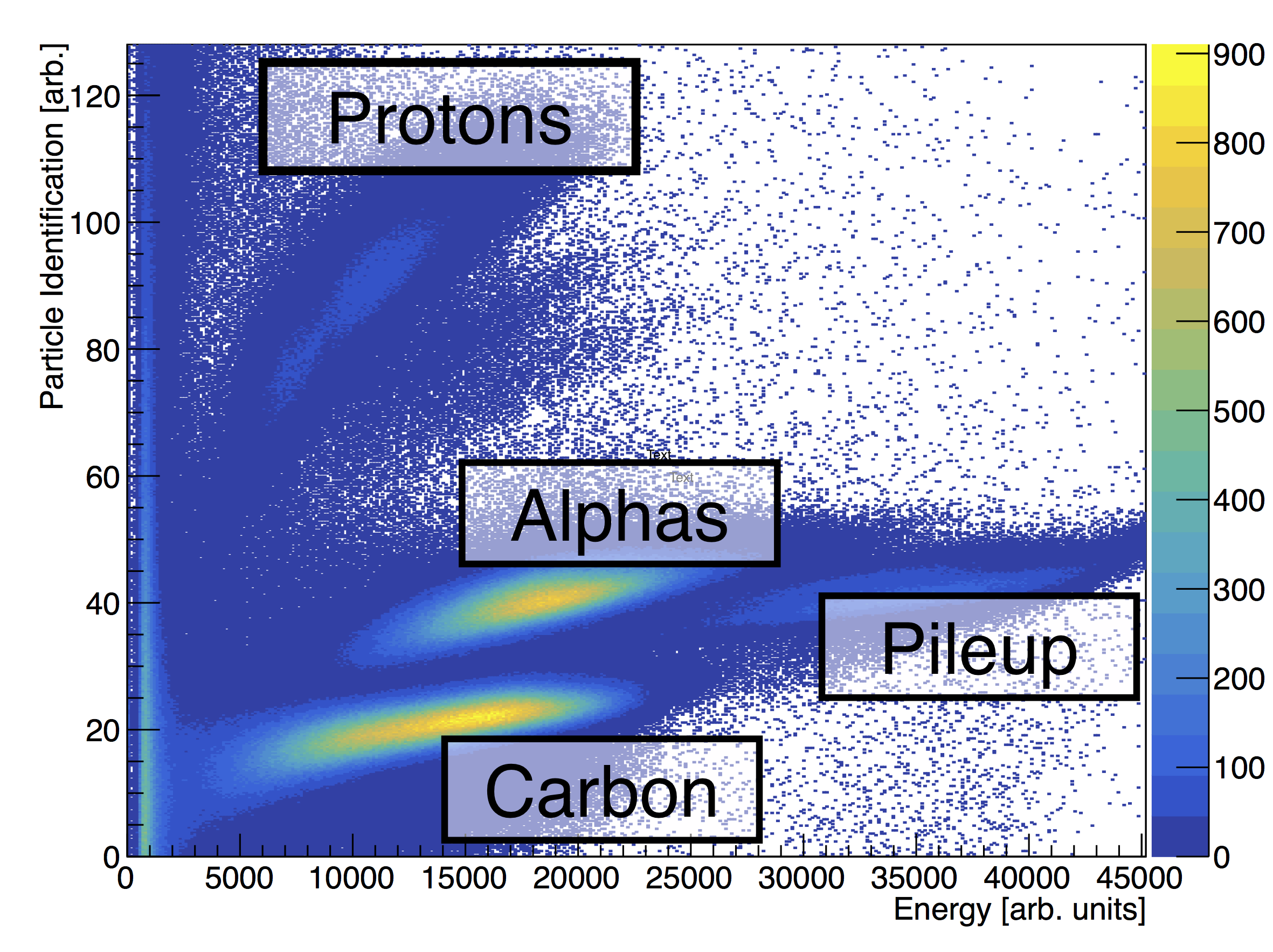}}
\caption{The particle identification (PID) parameter determined from waveform fitting plotted against the energy in the CsI(Tl) detectors with the major loci indicated.}
\label{fig:PID}
\end{figure}

States of interest were populated through the unsafe Coulomb excitation of $^{86}$Kr in inverse kinematics. A beam of $^{86}$Kr ions was produced by the TRIUMF offline ion source~\cite{ref:Jayamanna_08}, accelerated to 256.7~MeV by the TRIUMF-ISAC accelerator chain and delivered to the TRIUMF-ISAC gamma-ray escape-suppressed spectrometer (TIGRESS) facility~\cite{ref:Hackman_14}. For the present measurement, TIGRESS contained eleven 32-fold segmented HPGe clover detectors, with three located at $45^\circ$, five at $90^\circ$ and three at $135^\circ$ relative to the beam axis. TIGRESS surrounded the TIGRESS integrated plunger (TIP)~\cite{ref:Voss_14} setup which consisted of a wall of $24$ $16\times14$~mm$^2$, $2$-mm thick CsI(Tl) detectors located $51.7$~mm downstream of the target. The beam was impinged onto a 2.165-$\mu$m (0.5-mg/cm$^2$) thick amorphous carbon target, backed by a 14.917-$\mu$m (28.8-mg/cm$^2$) thick gold foil to stop the beam-like particles~\cite{ref:Green_14}. A beam intensity of approximately 100~ppA was maintained for 30 hours. Data were acquired using the TIGRESS digital data acquisition system~\cite{ref:Martin_08} with a particle-$\gamma$ coincidence condition required. Detector waveforms were collected for both the HPGe and CsI(Tl) detectors, as well as coincident accelerator-RF waveforms. A similar experimental configuration using TIP, TIGRESS and the same target used in the present work is discussed in Ref.~\cite{ref:Voss_17}.

\begin{figure}
\centerline{\includegraphics[width=\linewidth]{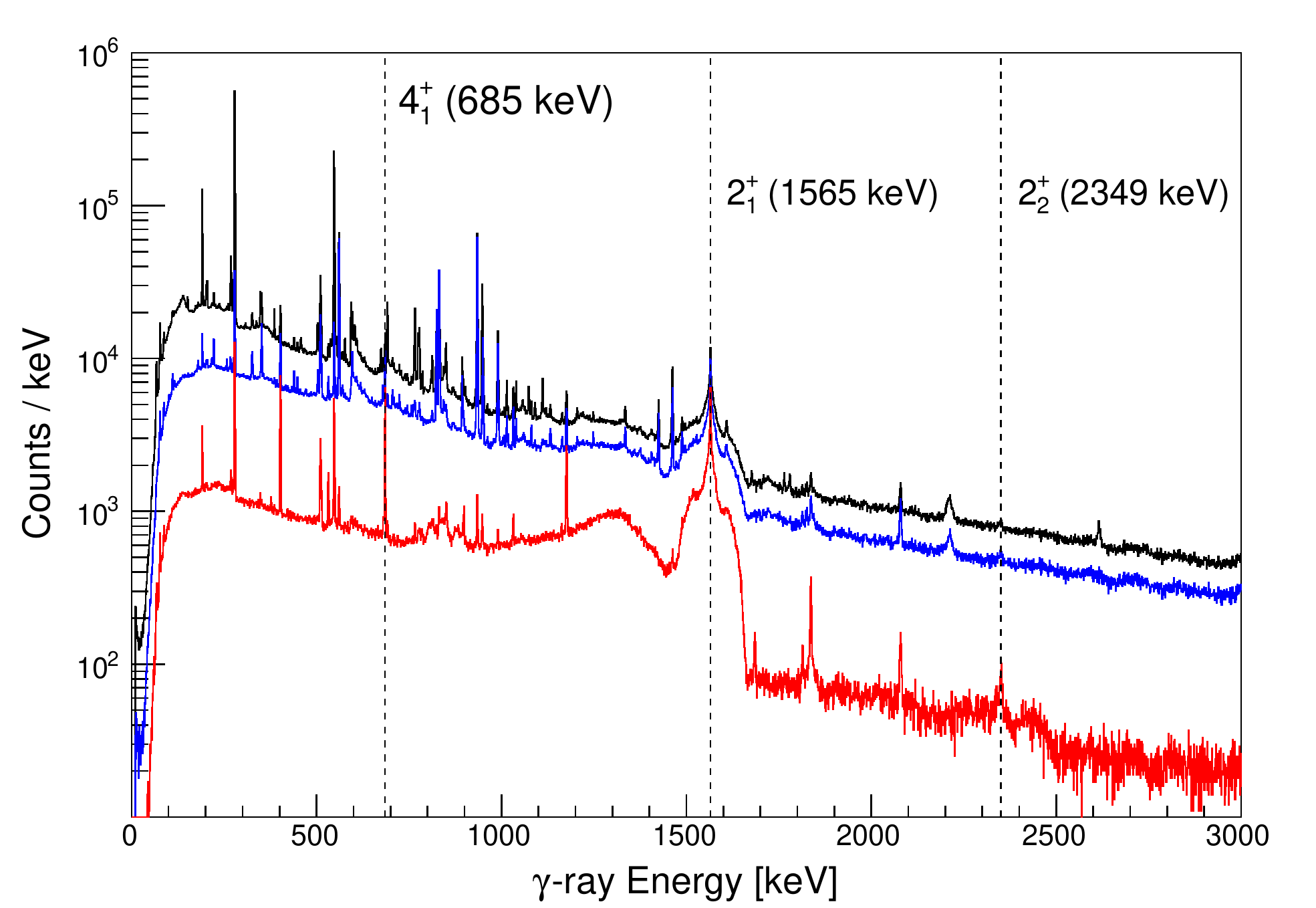}}
\caption{Black: Addback $\gamma$-ray spectrum. Blue: Requiring a good time coincidence with a CsI(Tl) event and an RF beam bucket. Red: Requiring the coincident CsI(Tl) event that has a PID parameter consistent with the a carbon detection. $\gamma$-ray peaks arising from the decays of states of interest are indicated. Note that $\gamma$-rays originating from the $3^-$ state are not resolvable from the those from the $2^+_1$ state in this figure.} 
\label{fig:Gammas}
\end{figure}

\begin{figure}
\centerline{\includegraphics[width=.9\linewidth]{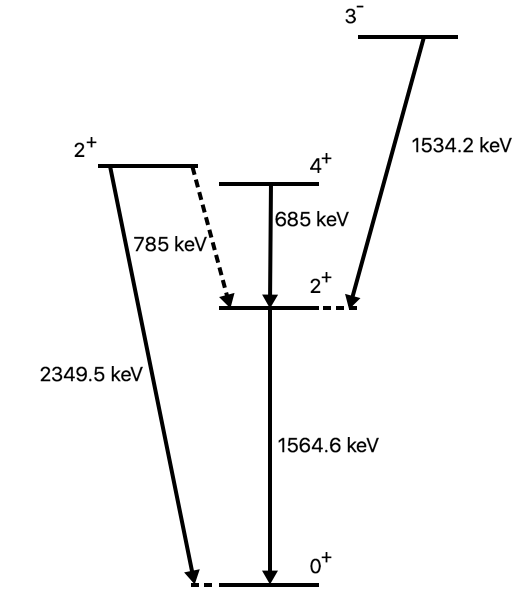}}
\caption{Reduced level scheme indicating the states of interest in the present work. Dashed transitions indicate $\gamma$ rays that were not observed but were incorporated into the lineshape analysis.}
\label{fig:levelscheme}
\end{figure}

\section{Analysis}
Data were unpacked and analyzed using the in-house GRSISort analysis software~\cite{ref:GRSISort}, built on the ROOT framework~\cite{ref:ROOT}. Events were constructed on the basis of a trigger ID, provided by the TIGRESS acquisition system. Precise relative timing information was extracted through the fitting of the CsI(Tl), HPGe and RF waveforms. Particle-$\gamma$ pairs were constructed on the basis of this timing information which were themselves coincident with an RF pulse and, thus, a beam bucket from the accelerator chain. Waveform fitting also allowed for particle identification (PID) on the basis of the well-understood response of CsI(Tl) (see, e.g. Ref.~\cite{ref:Voss_14}). A PID plot for the innermost four CsI(Tl) detectors is shown in Fig.~\ref{fig:PID}.

\begin{figure}
\centerline{\includegraphics[width=.6\linewidth]{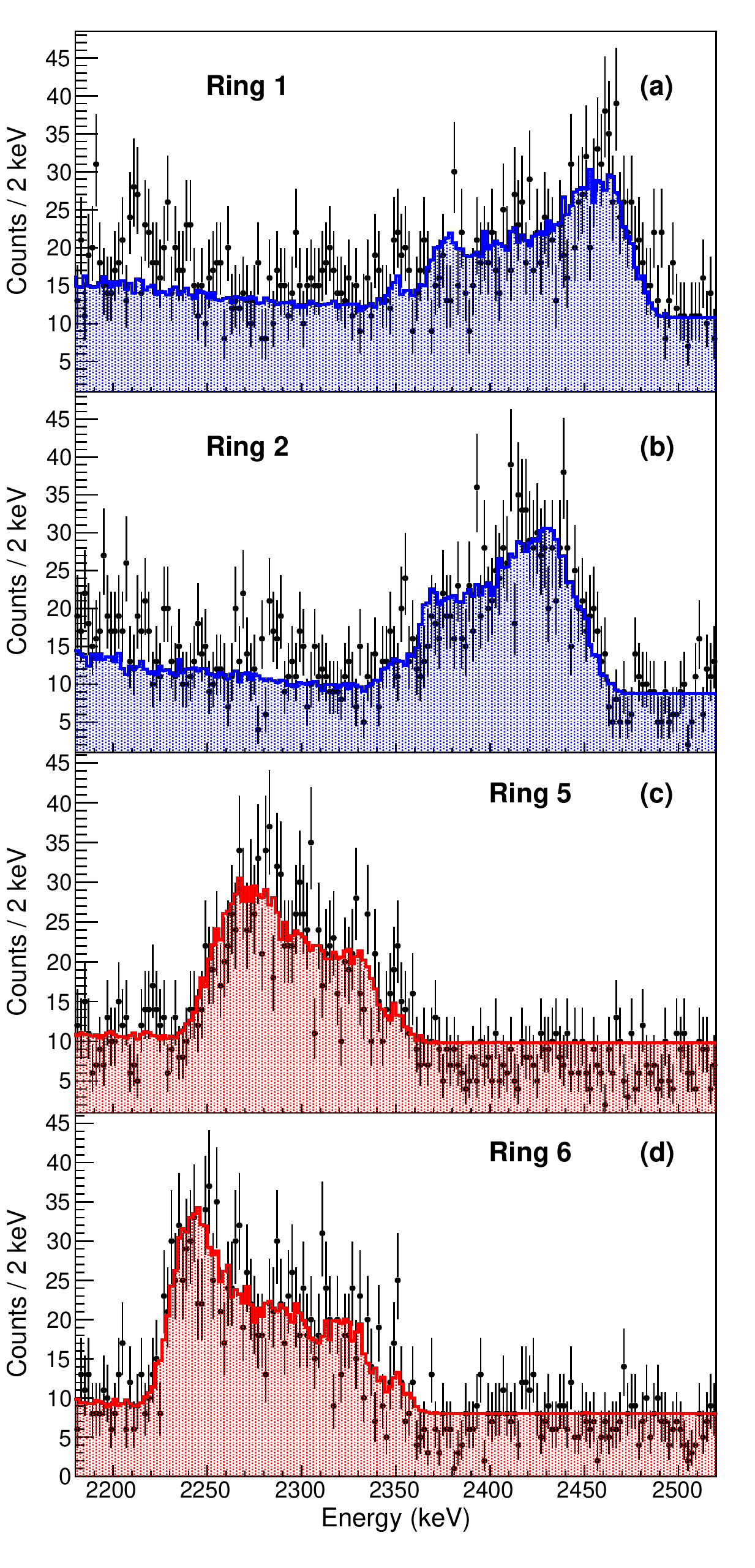}}
\caption{Simulated lineshape (filled) plotted along with the experimental data (data points) for rings 1, 2, 5 and 6 (a-d) for a simulated lifetime, $\tau(2^+_2)=260$~fs.}
\label{fig:Ring_6_2_2}
\end{figure}

\begin{figure}
\centerline{\includegraphics[width=\linewidth]{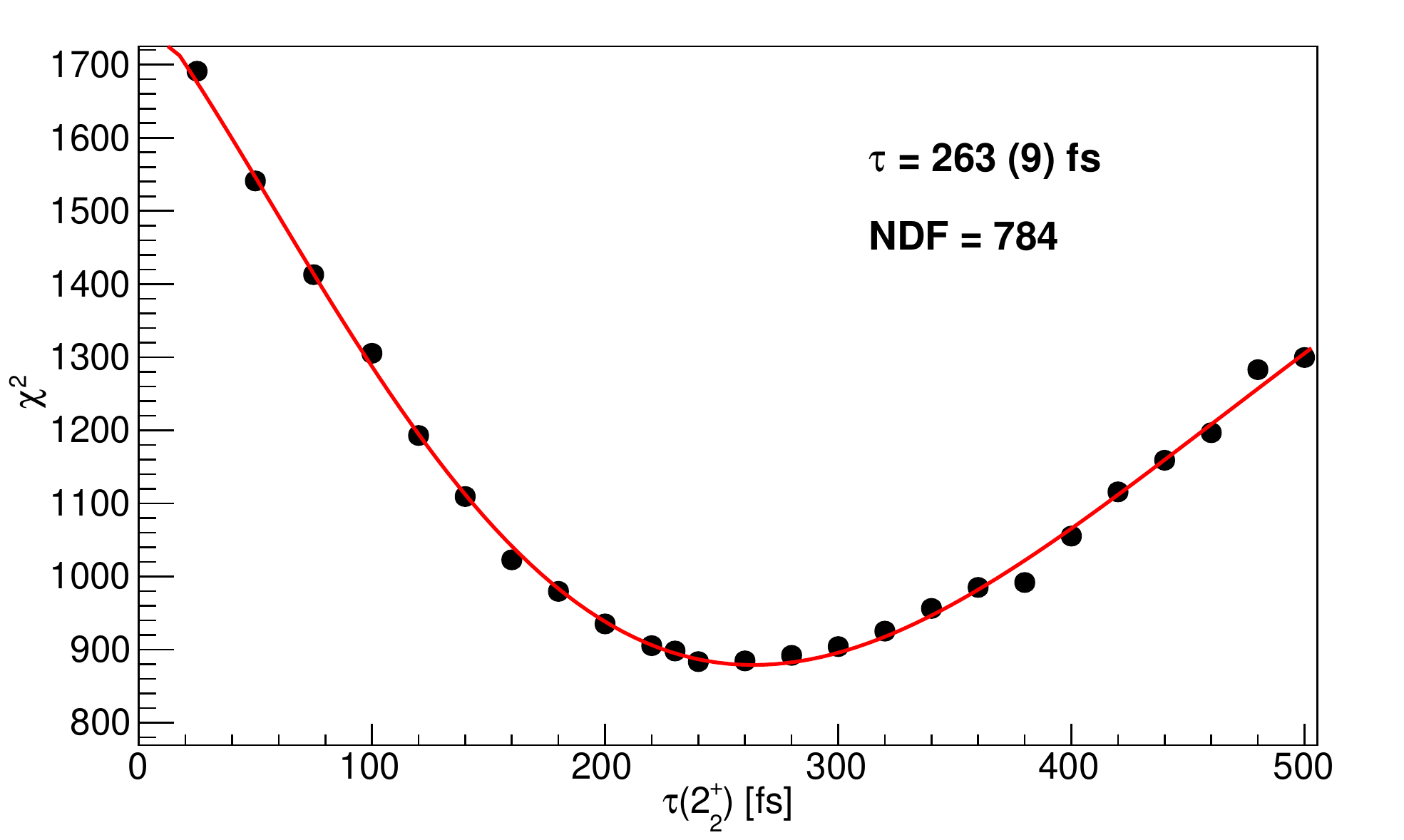}}
\caption{$\chi^2$ plotted against the simulated lifetime of the $2^+_2$ state for all four sensitive rings (see Fig.~\ref{fig:Ring_6_2_2}). As indicated, the $\chi^2$ minumum is found to correspond to $\tau=263$~(9)~fs. Once systematic uncertainties are incorporated the final value becomes $\tau=263$~(9)~(19)~fs.}
\end{figure}

In order to maximize peak-to-background, HPGe signals were added back on the basis of both the time of the detection event and the sub-clover position provided by the segmented detector anodes. The added-back data were then time gated on the basis of both an RF beam bucket and a time-coincident CsI(Tl) detector hit. Finally, a PID gate was applied, requiring the coincident CsI(Tl) event to be consistent with a carbon recoil. The $\gamma$-ray gating process is shown in Fig.~\ref{fig:Gammas}. The consequence of this series of filters was a very clean $\gamma$-ray spectrum. Some time-random $\beta$ decay lines remain in the spectrum, however these do not interfere with the DSAM lineshape analysis. The only discernible background arising from in-flight decays (i.e. not the result of time-random $\beta$ decays) were weak $\gamma$-ray lines consistent with a single-proton transfer reaction into $^{87}$Rb, the isotonic neighbour of $^{86}$Kr. Where resolvable these weak lines were fitted with a simple Gaussian and included in fitting of the Monte-Carlo simulations. No strong background lines were observed to interfere with the decays which will be discussed.

Observed in the data were four transitions from the decay of excited states of $^{86}$Kr: $2^+_1\rightarrow0^+_1$ ($E_\gamma = 1564.67$~keV), $4^+_1\rightarrow2^+_1$ ($E_\gamma = 685.35$~keV), $2^+_2\rightarrow0^+_1$ ($E_\gamma=2349.47$~keV) and $3^-_{(2)}\rightarrow2^+_1$ ($E_\gamma=1534.24$~keV). Of these transitions, only the $4^+_1\rightarrow2^+_1$ had no in-flight lineshape component, consistent with its previously determined halflife of 3.1~ns. The remaining three transitions were thus suitable for a Doppler-shift attenuation method (DSAM) analysis. A reduced level scheme showing the states observed in the present work is given in Fig.~\ref{fig:levelscheme}.

The techniques used to analyze the DSAM spectra are outlined in detail in Ref.~\cite{ref:Chester_18} and Ref.~\cite{ref:Williams_17}. High-statistics {\small GEANT4} simulations were performed using a detailed model of the TIGRESS and TIP spectrometers. For the purposes of the DSAM analysis, only the inner eight TIP CsI(Tl) detectors were used - for the remaining detectors the particle-identification plot (see Fig.~\ref{fig:PID}) did not give good separation for the carbon nuclei, and losses due to thresholds could not be consistently accounted for. The simulated {\small GEANT4} spectra were smeared according to the observed HPGe detector resolution prior to comparisons with the experimental data. For comparison with simulations, TIGRESS was separated into six rings corresponding to the downstream- and upstream-most rings of the $45^\circ$, $90^\circ$ and $135^\circ$ clovers, with ring 1 being the most downstream and ring 6 the most upstream. No sensitivity to lifetimes was found for the rings about $90^\circ$. Angular correlation effects were not found to be important and were not included in the further analysis.

\subsection{The $2^+_2\rightarrow0^+_1$ transition}

The $2^+_2\rightarrow0^+_1$ ($E_\gamma=2349$~keV) transition was well separated from the other lines and could thus be analyzed independently. No feeding transition to the $2^+_1$ state was observed ($E_\gamma=749$~keV), however this would not be expected to be visible above background. Simulated spectra were constructed from $5\cdot10^6$ CsI(Tl)-HPGe coincidences. The spectra were then fit to the experimental data for all four sensitive rings (1, 2, 5 \& 6) simultaneously. The parameters in the fits were a scaling parameter, a zeroth-order polynomial background which was constrained (but not fixed) prior to the fit, and an $E_\gamma$-shift parameter to account for any low-level mismatch between the simulated and experimental energies - in particular to allow for binning effects. Due to the low level of statistics for this transition, the use of a maximum-likelihood method was essential, as described in Ref.~\cite{ref:Chester_18}. The simulated lineshape corresponding to the maximized log-likelihood is shown in Fig.~\ref{fig:Ring_6_2_2}. To account for the fact that the reduced $\chi_{\text{min}}^2/\nu = 1.12 > 1$, the statistical uncertainties were inflated by a factor of $\sqrt[]{\chi^2_{min}/\nu}=1.06$ following the prescription of Ref.~\cite{ref:Patrignani_16}, yielding a final lifetime of 263~fs with $\delta\tau_\text{stat}=9$~fs. 

\subsection{The $2^+_1\rightarrow0^+_1$ and $3^-_{(2)}\rightarrow2^+_1$ transitions}

The $2^+_1\rightarrow0^+_1$ (1564~keV) and $3^-_{(2)}\rightarrow2^+_1$ (1534~keV) transitions lie too close in energy for their respective lineshapes to be fully disentangled. Indeed, in the downstream (rings 1 and 2) data, the $3^-_{(2)}\rightarrow2^+_1$ transition could not be visually identified. Consequently, and because the $3^-_{(2)}\rightarrow2^+_1$ transition feeds the $2^+_1$ state, the two transitions had to be analyzed simultaneously. No indication for any competing branch in the decay of the $3^-_{(2)}$ state was observed, nor is there any indication of such a decay in the literature. The branch was therefore assumed to be 100~\% in the present work. This is consistent with the literature $B(E3;3^-\rightarrow0^+)$ value extracted from proton scattering~\cite{ref:Kibedi_02,ref:Arora_74,ref:Matsuki_78} which yields a partial $E3$ lifetime of approximately 120~ps and a ground-state branch of less than 0.1\%.

\begin{figure}
\centerline{\includegraphics[width=\linewidth]{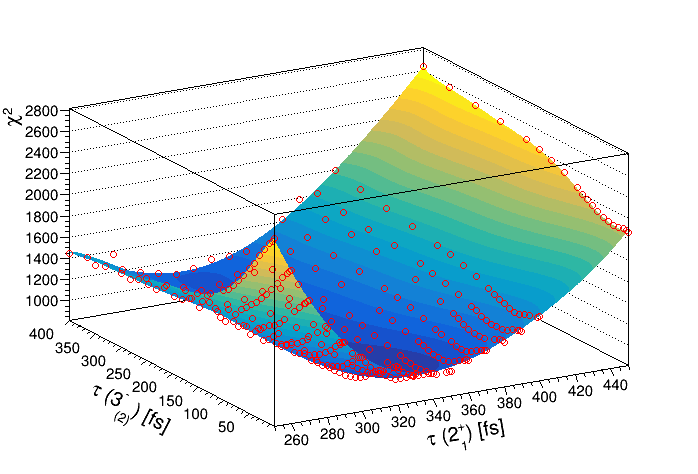}}
\caption{The $\chi^2$ surface fitted to the minimized $\chi^2$ values (red circles) used to determine the lifetime minima for the overlapping $2^+_1$ and $3^-_{(2)}$ lineshapes.}
\label{fig:ChisqSurf}
\end{figure}

\begin{figure}
\centerline{\includegraphics[width=.7\linewidth]{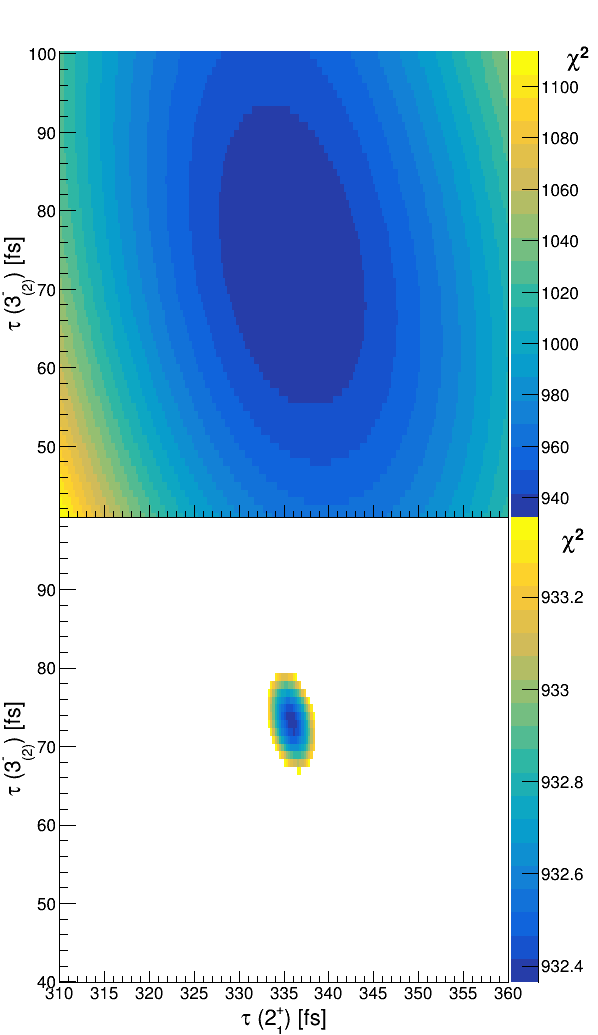}}
\caption{Top: $\chi^2$ surface as determined from the fit shown in Fig.~\ref{fig:ChisqSurf} in the region of $\chi^2_\text{min}$. Bottom: The $\chi^2\leq\chi^2_{\text{min}}+1$ region from which the minimum and associated statistical uncertainties were determined.}
\label{fig:Surf_1Sig}
\end{figure}

\begin{figure*}
\centerline{\includegraphics[width=.8\textwidth]{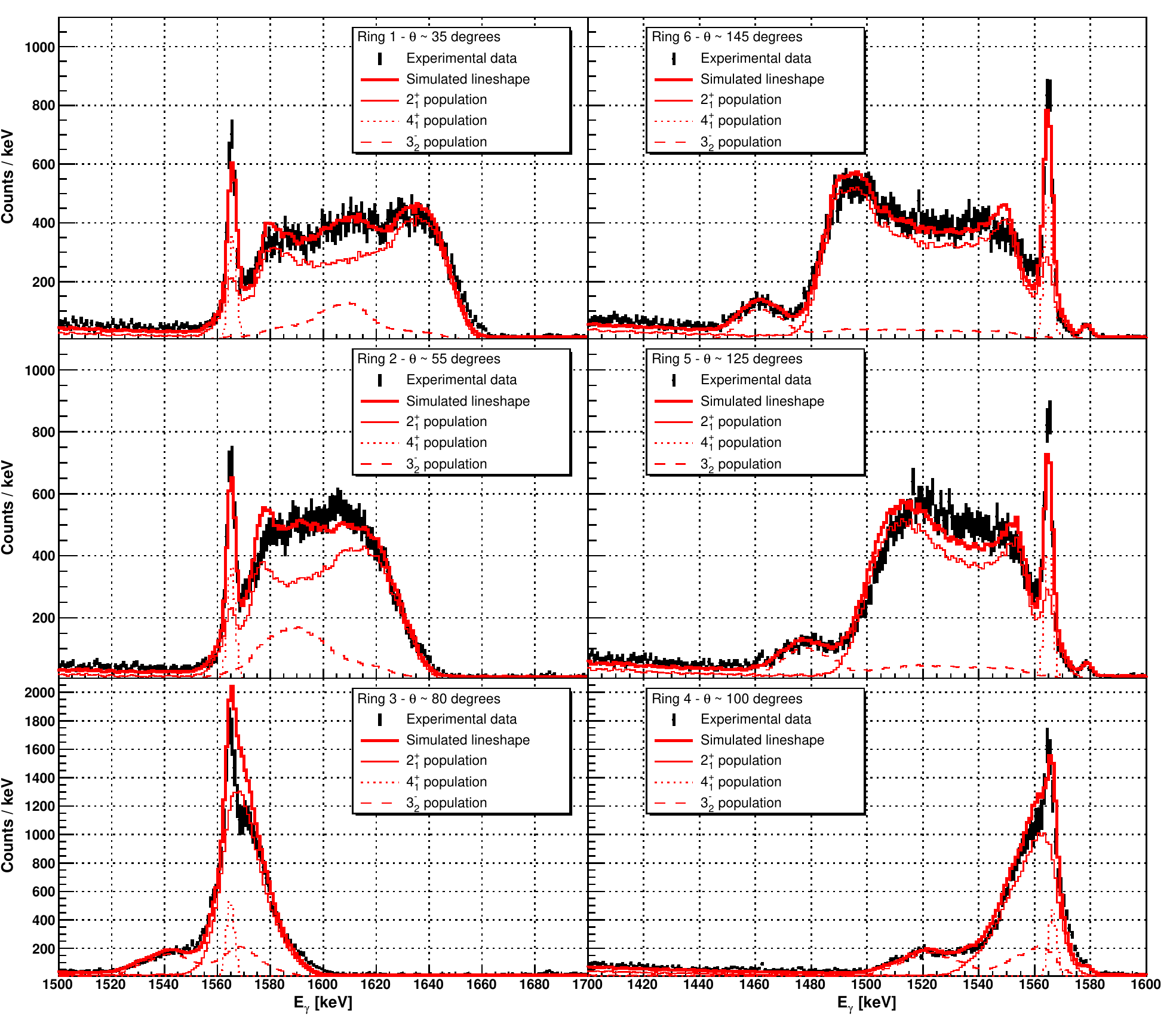}}
\caption{Fitted lineshape (thick solid, red) corresponding to the minimized $\chi^2$ overlayed on the experimental data (black points) for the six rings in TIGRESS. Also shown are the primary contributions to the lineshape arising from direct population of the $2^+_1$ state (thin solid, red), population via the $4^+_1$ state (dotted, red) and via the $3^-_{(2)}$ state (dashed, red). Details of the parameters varied in the fits are provided in the text. Rings 3 and 4 were insensitive to the state lifetimes and were not included in the $\chi^2$ surface shown in Fig.~\ref{fig:ChisqSurf}.}
\label{fig:lineshapes}
\end{figure*}

Because the two lineshapes overlapped it was necessary for all rings to be analyzed simultaneously, with the backward rings being particularly sensitive to short $3^-_{(2)}$ lifetimes and the forward rings placing limits on long lifetimes. Simulated spectra were again generated with $5\cdot10^6$ CsI(Tl)-$\gamma$ coincident events, with the feeding and subsequent decay of the $2^+_1$ state included in the simulation of the $3^-$ decay (see Ref.~\cite{ref:Williams_17}). 340 simulated data sets were therefore created for the $3^-$ decay, corresponding to seventeen potential $2^+_1$ lifetimes and twenty potential $3^-_{(2)}$ lifetimes, with a further seventeen simulated datasets corresponding to direct population of the $2^+_1$ state. Feeding of the $2^+_1$ state from the stopped ($t_{1/2}$=3.1~ns) $4^+_1$ state was also simulated. Feeding from the $2^+_2$ state was included but was found to have very little effect on the final result. 

The simulated data were then fit to the experimental spectra, with all four rings fitted simultaneously. For each ring, the following parameters were independently varied to achieve the best fit: a scaling parameter for the $2^+_1$ direct feeding, a scaling parameter for the $3^+_{(2)}$ decay (including the resultant decay of the $2^+_1$ state), a zeroth order polynomial background and an $E_\gamma$-shift parameter to allow for low-level mismatch between simulated and experimental energies. The $4^+$ feeding contribution was fixed for all rings based on a fit to the intensity of the $4^+_1\rightarrow2^+_1$ decay line.

A $\chi^2$ surface was fit to the resultant $\chi^2$ values, as shown in Fig.~\ref{fig:ChisqSurf}. From this, the $\chi^2_{min} \leq \chi^2 \leq \chi^2_{min} + 1$ range could then be extracted to determine the resultant state lifetimes and their associated uncertainties. Figure~\ref{fig:Surf_1Sig} shows the total and $1\sigma$ surface in the vicinity of the minimum with the fitted lineshapes shown in Fig.~\ref{fig:lineshapes}, along with the major contributions to the fit. The resultant lifetimes are $\tau(2^+_1)=336\pm3$~fs and $\tau(3^-_{(2)})=73\pm5$~fs. To account for the fact that the reduced $\chi_{\text{min}}^2/\nu = 1.6 > 1$, the statistical uncertainties were inflated by a factor of $\sqrt[]{\chi^2_{min}/\nu}=1.3$ following the prescription of Ref.~\cite{ref:Patrignani_16}, yielding final lifetimes of $\tau(2^+_1)=336$~fs with $\delta\tau_\text{stat}(2^+_1)=4$~fs and $\tau(3^-_{(2)})=73$~fs with $\delta\tau_\text{stat}(3^-_{(2)})=6$~fs. 

\begin{figure*}
\centerline{\includegraphics[width=.8\linewidth]{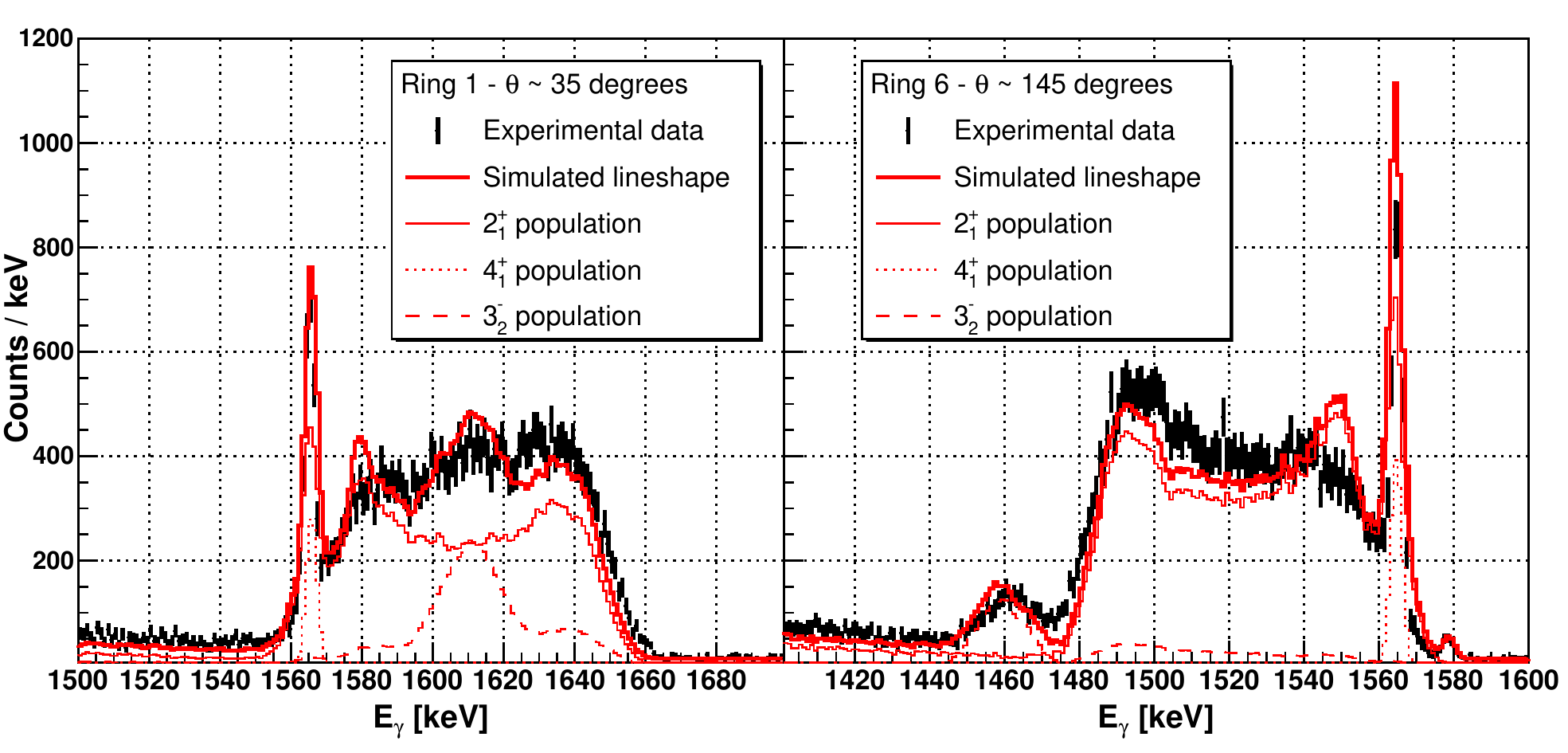}}
\caption{Lineshapes for a $2^+_1$ state lifetime of 450~fs (Ref.~\cite{ref:Mertzimekis_01}: $\tau(2^+_1)=440 (25)$ fs) for the downstream- and upstream-most rings. A $3^-_{(2)}$ state lifetime of $\tau(3^-_{(2)})=10$~fs is used, corresponding to the approximate $\chi^2$ minimum in Fig.~\ref{fig:ChisqSurf} at $\tau(2^+_1)=450$~fs. The poorer quality of the fit for both the $2^+_1$ and $3^-$ components is clear in comparison to that of Fig.~\ref{fig:lineshapes}.}
\label{fig:lineshape_450fs}
\end{figure*}
\begin{figure}
\centerline{\includegraphics[width=1.05\linewidth]{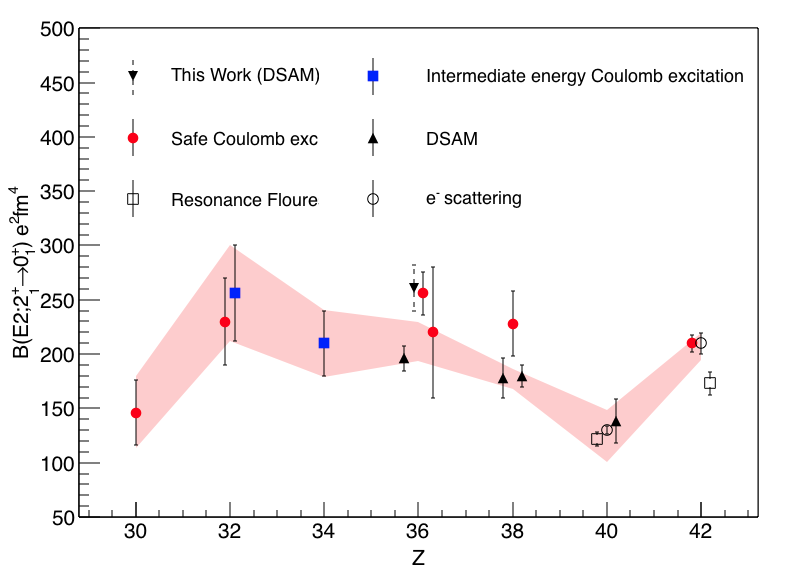}}
\caption{$B(E2;2^+_1\rightarrow0^+_1)$ value systematics for the $N=50$ isotones. The uncertainties on the present result are dominated by systematics (dashed lines) but agree well with previous Coulomb excitation data. The evaluated data range is indicated by the shaded band. Experimental data taken from Refs.~\cite{ref:deWalle_07,ref:Padilla-Rodal_05,ref:Gade_10,ref:Mertzimekis_01,ref:Cheng-Lie_81,ref:Cartwright_81,ref:Kucharska_88,ref:Christensen_73,ref:Kumbartzki_14,ref:Heisenberg_84,ref:Metzger_72,ref:Sahota_93,ref:Raman_87,ref:Metzger_77,ref:Milliman_87}.} 
\label{fig:N50_BE2}
\end{figure}

\subsection{Systematic uncertainties}

Uncertainties on the present measurement are dominated by systematic effects. The primary systematic uncertainties in the present measurement arise from uncertainties in the stopping powers for krypton in amorphous carbon and gold. The fast (large $v/c$) component to the lineshapes was found to be sensitive to the stopping in carbon and was varied during the simulation procedure. All results thus far correspond to a stopping power of 60\% of that included by default within the {\small GEANT4} libraries which was found to best reproduce the data. Full simulations were performed for 40\%, 50\%, 60\%, 70\% and 80\% of the nominal carbon stopping powers in {\small GEANT4}. Based on simulations performed with these stopping powers systematic uncertainties of $\delta\tau(2^+_1)=7$~fs and $\delta\tau(3^-_{(2)})=4$~fs were estimated. A similar analysis was performed for the lifetime of the $2^+_2$ state, yielding an uncertainty from the carbon stopping powers of $\delta\tau(2^+_2)=5$~fs. No clear sensitivity was discernible in the fitting to the stopping of krypton in gold. 
We attribute a 5\% additional systematic uncertainty to the final results from potential discrepancies in gold stopping powers. A small systematic uncertainty (2~fs) is also associated with the choice of polynomial used to fit the $\chi^2$ surface in Fig.~\ref{fig:ChisqSurf}. We find a small sensitivity to feeding from the $2^+_2$ state, from which we attribute a 1~fs systematic uncertainty to the $2^+_1$ and $3^-$ lifetimes. Finally, it is found that for the $3^-_{(2)}$ state there is a discrepancy between the lifetime extracted using rings one and six ($35^\circ$ and $145^\circ$, 96~(10)~fs), and that extracted using rings two and five ($55^\circ$ and $125^\circ$, 53~(11)~fs). We attribute a $20$~fs systematic uncertainty, accordingly. The equivalent discrepancy for the $2^+_1$ and $2^+_2$ states is found to be approximately 2~fs. We therefore quote final systematic uncertainties of $\delta\tau_\text{(sys.)}(2^+_1)=20$~fs, $\delta\tau_{\text{(sys.)}}(2^+_2)=19$~fs and $\delta\tau_\text{(sys.)}(3^-_{(2)})=32$~fs, giving results of $336\pm4\text{(stat.)}\pm20\text{(sys.)}$~fs for the $2^+_1$ state, $263\pm9\text{(stat.)}\pm19\text{(sys.)}$~fs for the $2^+_2$ state, and $73\pm6\text{(stat.)}\pm32\text{(sys.)}$~fs for the $3^-_2$ state. 

\section{Discussion}

\begin{table}
\caption{Lifetimes and reduced transition probabilities as determined in the present work compared to literature values, where available. The branching ratio of the $2^+_2$ state was taken from Ref.~\cite{ref:ENSDF}. The quoted $B(E1)$ for the $3^-\rightarrow2^+$ transition assumes a pure $E1$. No other decay branches for the $3^-_{(2)}$ state are known, nor were any observed in the present work. Here, the first quoted uncertainty corresponds to statistics, the second to systematics.}
\label{tab:Results}
\begin{ruledtabular}
\begin{tabular}{llllll}
 & \multicolumn{2}{c}{$\tau$ (fs)} & \multicolumn{2}{c}{$B(E\lambda)$ e$^2$fm$^{2\lambda}$} & \\ 
 \hline \\[-6pt]
Transition & This Work & Lit. & This Work & Lit. & Ref. \\
 \hline \\[-6pt]
$2^+_1\rightarrow0^+_1$ & 336 (4)(20) & 444 (25) & 259 (3)(16) & 196 (11) & \cite{ref:Mertzimekis_01} \\
& & 341 (27) & & 256 (20) & \cite{ref:Cheng-Lie_81} \\
& & 396 (108) & & 220 (60) & \cite{ref:Cartwright_81} \\
$3^-_{(2)}\rightarrow2^+_1$ & 73 (6)(32) & & 0.0024 (2)(11) & & \\
$2^+_2\rightarrow0^+_1$ & 263 (9)(19) & & 31 (1)(2) & & \\
$2^+_2\rightarrow2^+_1$ & 263 (9)(19) & & $<$3000 & & \\
\hline \\[-6pt]
Transition & \multicolumn{2}{c}{$\tau$ (fs)} & \multicolumn{2}{c}{$B(M1)$~$\mu^2_N$} \\
\hline \\[-6pt]
$2^+_2\rightarrow2^+_1$ & \multicolumn{2}{c}{263 (9)(19)} & \multicolumn{2}{c}{$<$ 0.13} \\
\end{tabular}
\end{ruledtabular}
\vspace{-10pt}
\end{table}

The present results deviate from those determined in Ref.~\cite{ref:Mertzimekis_01} by more than three standard deviations. We note, however, that the measurement reported in Ref.~\cite{ref:Mertzimekis_01} was unable to resolve the $3^-$ state observed in the present work. The failure to include this contribution in the lifetime determination of the present work - even at relatively low levels - would result in the extraction of a longer lifetime from the analysis. This highlights the importance of using HPGe arrays with detectors at both forward and backward angles in DSAM analyses. In order to demonstrate to the reader the incompatibility of the lifetime quoted in Ref.~\cite{ref:Mertzimekis_01} with the present data, Fig.~\ref{fig:lineshape_450fs} shows simulated lineshapes corresponding to a lifetime of 450~fs. The lifetime extracted in the present work is in good agreement with both the Coulomb excitation works of Ref.~\cite{ref:Cheng-Lie_81} and Ref.~\cite{ref:Cartwright_81}. Excluding the DSAM result of Ref.~\cite{ref:Mertzimekis_01} for the above stated reasons, we determine a new weighted average lifetime, $\tau_{2^+_1}=339\pm 16$~fs.

Figure~\ref{fig:N50_BE2} shows $B(E2;2^+_1\rightarrow0^+_1)$ values for the $N=50$ isotones, where known. Clearly, the present result and its associated conclusions represents a significant deviation from the accepted values, with $^{86}$Kr now becoming one of the more collective of the $N=50$ nuclides. In light of the present result, a remeasurement of the $B(E2;2^+_1\rightarrow0^+_1)$ value in $^{88}$Sr would be of interest in order to confirm behaviour approaching $Z=40$.

\section{Conclusions}

We have remeasured the lifetime of the first-excited state in $^{86}$Kr using the Doppler-shift attentuation method (DSAM) following population in unsafe Coulomb excitation. Our result agrees with previous Coulomb-excitation measurements and disagrees with a previous DSAM measurement at the $3\sigma$ level. We hypothesize that this discrepancy may arise from the failure of the previous measurement to resolve a feeding state with a lineshape that overlaps with the state of interest. We were further able to extract lifetimes for the $3^-_{(2)}$ and $2^+_2$ states and determine transition strengths accordingly. Our new data indicate a more precipitous reduction in $B(E2)$ strength in the $N=50$ isotones approaching the $Z=40$ sub-shell closure than was previously thought to occur. This new lifetime may also affect the conclusions of g-factor measurements which need to account for the lifetimes of the states of interest (e.g. Ref.~\cite{ref:Mertzimekis_01} and Ref.~\cite{ref:Kumbartzki_14}) in the transient-field technique.

\section{Acknowledgements}

The authors would like to thank the TRIUMF beam delivery group for their efforts in providing high-quality beams. This work has been supported by the Natural Sciences and Engineering Research Council of Canada (NSERC), The Canada Foundation for Innovation and the British Columbia Knowledge Development Fund. TRIUMF receives federal funding via a contribution agreement through the National Research Council of Canada. The work at LLNL is under contract DE-AC52-07NA27344.

\bibliographystyle{unsrt}
\bibliography{kr86}

\end{document}